\documentclass[pra,onecolumn,showpacs,eqsecnum,preprintnumbers,amsmath,amssymb,superscriptaddress, eqsecnum]{revtex4}

\usepackage{bm}
\usepackage{graphicx}
\usepackage{amsbsy} 
\usepackage{amsmath}
\usepackage{amsfonts}
\usepackage{amsthm}
\usepackage{bm}
\usepackage{graphicx}
\usepackage{amsbsy}
\usepackage{amsmath}
\usepackage{amsfonts}
\usepackage{amsthm}
\usepackage{color}
\usepackage{mathrsfs}

\begin{document}

\theoremstyle{plain}
\newtheorem{theorem}{Theorem}
\newtheorem{lemma}[theorem]{Lemma}
\newtheorem{corollary}[theorem]{Corollary}
\newtheorem{conjecture}[theorem]{Conjecture}
\newtheorem{proposition}[theorem]{Proposition}

\theoremstyle{definition}
\newtheorem{definition}{Definition}

\theoremstyle{remark}
\newtheorem*{remark}{Remark}
\newtheorem{example}{Example}

\title{Time reversal frameness and superselection}

\author{Gilad Gour}\email{gour@math.ucalgary.ca}
\affiliation{Institute for Quantum Information Science, University of Calgary, Calgary, Alberta T2N 1N4, Canada}
\affiliation{Department of Mathematics and Statistics, University of Calgary, Calgary, Alberta T2N 1N4, Canada} 
\author{Barry C. Sanders}
\affiliation{Institute for Quantum Information Science, University of Calgary, Calgary, Alberta T2N 1N4, Canada}
\author{Peter S. Turner}
\affiliation{Institute for Quantum Information Science, University of Calgary, Calgary, Alberta T2N 1N4, Canada}
\affiliation{Department of Physics, Graduate School of Science, University of Tokyo, Tokyo 113-0033, Japan}

\date{\today}

\begin{abstract}
We show that appropriate superpositions of motional states are a reference frame resource
that enables breaking of time -reversal superselection so that two
parties lacking knowledge about the other's direction of time can still communicate. 
We identify the time-reversal reference frame resource states and determine
the corresponding frameness monotone, which connects time-reversal frameness to entanglement. 
In contradistinction to other studies of reference frame quantum resources, this
is the first analysis that involves an antiunitary rather than unitary representation.
\end{abstract}

\maketitle

\section{Introduction}

Superselection in quantum mechanics~\cite{AS67} is the principle
that certain types of coherent superpositions are forbidden,
and operations on Hilbert space must be invariant under group transformations that would allow violation of the superselection principle.
At times, superselection has been touted as a fundamental principle to prevent such counterintuitive
phenomena such as superpositions of different charges that are otherwise permissible in quantum mechanics.
More recently superselection has been given an informational interpretation that is associated with 
lack of a reference frame~\cite{RS01,BRS07} such as orientation, chirality, or phase~\cite{GS08}.
In the cases studied so far, the superselection is separated from coherence by ignorance or knowledge of which unitary transformation
to apply in a particular frame, and this knowledge is provided by a reference frame that can be quantum
or classical in nature. In the quantum case, the reference frame is consumed through its use to 
measure quantities that would otherwise be forbidden by superselection~\cite{BRST06,BRST07}.

The relationship between reference frames and superselection bears similarity to 
entanglement theory and restrictions under local operations and classical communication (LOCC)~\cite{PV07}.
Moreover entanglement as a resource to break LOCC restrictions is related to reference
frames as a resource for violating superselection. The reference frame resource is known
as \emph{frameness}, and there are corresponding monotones analogous to entanglement monotones;
these frameness monotones are non-decreasing under $G$-invariant operations, where $G$
is the group associated with the super selection rule. 

Here we consider the problem of time reversal invariance. Consider two parties, Alice and Bob,
who wish to communicate but do not know which direction in time applies to the other party.
In an exotic setting, Alice and Bob could be both moving forwards or backwards in time,
or one could be moving forward and the other backwards in time.
In a more practical situation, Alice and Bob could be unsure whether their directions of
motion are co-aligned or counter-aligned, as time-reversal corresponds to motion reversal.

The lack of knowledge about direction of time yields a time-reversal (TR) superselection rule
that can be broken by Alice and Bob sharing a TR resource.
Whereas previous studies of resources to break superselection rules have all involved
invariance under unitary group actions, TR is novel in that the corresponding transformations
are antiunitary. Here we introduce the notion of a TR-invariant operation (TRIO) and 
corresponding TRIO equivalence classes, develop a resource theory for TR, establish
three types of TRIO frameness monotones analogous to the entanglement monotone
types `deterministic'~\cite{Nie99}, `ensemble'~\cite{JP99}, and `stochastic'~\cite{Vid99}, 
and we propose TR frameness distillation analogous to entanglement distillation~\cite{BBPS96}.

\section{Representation of time reversal}
\label{sec:representations}

Classical mechanics is TR-invariant: the discrete transformation of extended phase space,~$\mathsf{T}$,
that fixes positions and reverses both momenta and time
\begin{align}
	\mathsf{T}:\mathbb{R}^3\times\mathbb{R}^3\times\mathbb{R}
		\rightarrow\mathbb{R}^3\times\mathbb{R}^3\times\mathbb{R}:
	 (\bm{x},\bm{p},t) \mapsto (\bm{x},-\bm{p},-t)
\end{align}
does not change Hamilton's equations of motion for TR symmetric Hamiltonians.  Note that classical angular momentum,~$\bm{L} = \bm{x} \times \bm{p}$, is also reversed under this transformation.  It is clear that two successive time reversals give the identity transformation,~$\mathsf{T}^2=\mathsf{I}$.  Thus,~$\{\mathsf{T},\mathsf{I}\}\cong S_2$, which is the symmetric group of two elements.

In quantum mechanics, TR must be a transformation of Hilbert space: 
we denote a representation of~$S_2$ carried by bounded operators on Hilbert space by~$\Theta$
\begin{equation}
	\Theta:S_2 \rightarrow \mathcal{B}(\mathscr{H}):
	\left\{\begin{array}{c}
		\mathsf{T} \mapsto \hat\theta	\\
		\mathsf{I} \mapsto \openone
		\end{array}
	\right. .
\end{equation}
Following Wigner~\cite{Wig31}, if quantum mechanics is also to be TR-invariant, then the commutation relations $[\hat{\bm{x}},\hat{\bm{p}}]=i$ must be left invariant by~$\hat\theta$, which gives
\begin{equation}
\hat\theta[\hat{\bm{x}},\hat{\bm{p}}]\hat\theta^{-1}
= [\hat{\bm{x}},-\hat{\bm{p}}] 
= -\!i 
= \hat\theta i \hat\theta^{-1}.
\end{equation}
This implies that~$\hat\theta$ must be antilinear, equivalent to a linear operator composed with the complex conjugation operator.  Since complex conjugation must be performed with respect to some basis,~$\hat\theta$ is basis dependent and the antilinearity is given by
\begin{equation}\label{eq:alin}
\hat\theta \left( a|\psi\rangle + b|\phi\rangle \right)
= a^\ast \hat\theta|\psi\rangle + b^\ast \hat\theta|\phi\rangle, \quad a,b\in\mathbb{C},
\end{equation}
for $|\psi\rangle,|\phi\rangle$ in this basis.

Following Sakurai~\cite{Sak94} and  Landau and Lifshitz~\cite{::91} we choose as our basis states those of multiple orbital angular momenta
\begin{align}\label{eq:discretewavefunction}
\mathscr{H} = \mathrm{span}\{|\mu \ell m \rangle\}, \quad \langle \mu \ell m|\mu' \ell' m'\rangle = \delta_{\mu\mu'} \delta_{\ell\ell'} \delta_{mm'},
\end{align}
where~$\ell\in\mathbb{N}$ labels the angular momentum with 
\begin{equation}
	\text{spec}(\hat{\bm{L}}^2)=\{ \ell(\ell+1) \}, \;
	m\in\mathrm{spec}(\hat{L}_z)=\{ -\ell, -\ell+1,\cdots,\ell \},
\end{equation}
and~$\mu\in\mathbb{Z}^+$ is a multiplicity label distinguishing orthogonal states with the same angular momentum.  Interpreting $m$ as the projection of a classical total angular momentum vector onto the $z$-axis, and since classical angular momentum is reversed by TR, suggests that the TR operator~$\hat\theta$ should send states with quantum number $m$ to those with quantum number $-m$.
Including a state- (but not multiplicity-) dependent phase gives the TR action
\begin{align}\label{eq:thetajm}
\hat\theta|\mu\ell m\rangle = \text{e}^{\text{i}\theta_{\ell m}}|\mu\ell\,-\!m\rangle.
\end{align}

\begin{remark}
	Time reversal~$\Theta$ does not mix states of different~$\ell$;
	hence the SO(3) irreps labeled by~$\ell$ are intact under TR.
\end{remark}

The phase~$\theta_{\ell m}$ in Eq.~(\ref{eq:thetajm}) is not unique, 
although it must vary as~$\pm\pi m$, and is given differently by several authors.  
For example, Sakurai~\cite{Sak94} gives~$\theta_{\ell m}=\pi m$,
whereas Landau and Lifshitz~\cite{::91} employ
\begin{equation}
\label{eq:LLconvention}
	\theta_{\ell m}=\pi(\ell-m).
\end{equation}
Either convention is satisfactory in this section, but later we prefer the Landau and Lifshitz convention as it respects the Clebsch-Gordan decomposition for two particles, for which the choice~$\theta_{\ell m}=\pi(\ell-m)$ yields
\begin{align}\label{eq:CG}
\hat\theta \left( |\ell m\rangle |\ell' m'\rangle \right)
=& \hat\theta \sum_{L} (\ell m;\ell' m'|LM)|LM\rangle\nonumber \\
=& \sum_L (\ell m;\ell' m'|L M)^\ast \text{e}^{\text{i}\pi(L-M)} |L\;-\!M\rangle\nonumber \\
=& (-1)^{\ell+\ell'+M'}\sum_L (\ell -\!m;\ell' -\!m'|L M')|LM'\rangle\nonumber \\
=& \hat\theta|\ell m\rangle\otimes\hat\theta|\ell' m'\rangle
\end{align}
where we have used real Clebsch-Gordan coefficients, the dummy variable $M'=-m-m'$, and the identity
\begin{equation}
\label{eq:CGidentity}
(\ell m;\ell' m'|L M) = (-1)^{\ell+\ell'-L} (\ell\,-\!m;\ell'\,-\!m'|L\,-\!M).
\end{equation}
For the choice~$\theta_{\ell m}=\pi(\ell-m)$, we have
\begin{equation}
\label{eq:tensorproductaction}
	\hat\theta(\vert\psi\rangle \otimes \vert\phi\rangle)
		=\hat\theta\vert\psi\rangle \otimes \hat\theta\vert\phi\rangle,
\end{equation}
which defines TR for multiple systems.

Although the basis $\{|\mu l m\rangle\}$ is motivated by physics, for the purpose of quantifying the resources, we will use
a different basis that is mathematically much more natural to work with. That is, the {\it self-conjugate} basis. According to Lemma~1
in~\cite{Gar06}, for any conjugation like the time reversal operation $\hat\theta$, there exists an orthonormal basis $\{|e_n\rangle\}$, 
called the self-conjugate basis for which $\hat\theta|e_n\rangle=|e_n\rangle$. In the following lemma, we express this self-conjugate basis
in terms of the basis $\{|\mu l m\rangle\}$.

\begin{lemma}
	The self-conjugate basis $|e_n\rangle$, where $n$ stands for the set $\{\mu lm\pm\}$, is given by 
	\begin{align}
	\label{eq:eigenbasis}
		|e_{\mu\ell0+}\rangle := & \text{e}^{\text{i}\theta_{\ell 0}/2} |\mu\ell 0\rangle, \qquad \ell\geq 0, m=0, \nonumber \\
		|e_{\mu\ell m+}\rangle := & \frac{1}{\sqrt{2}}\left( |\mu\ell m\rangle 
			+ \text{e}^{\text{i}\theta_{\ell m}}|\mu\ell\,-\!m\rangle \right), \; 0<m\leq\ell\neq 0\;,\nonumber\\
		|e_{\mu\ell m-}\rangle := & \frac{i}{\sqrt{2}}\left( -|\mu\ell m\rangle 
			+ \text{e}^{\text{i}\theta_{\ell m}}|\mu\ell\,-\!m\rangle \right), \; 0<m\leq\ell\neq 0.
	\end{align}
\end{lemma}
\begin{proof}
Consider an arbitrary state expanded in the original basis~(\ref{eq:discretewavefunction}):
\begin{equation}
\label{eq:arbitrarystate}
	\vert\psi\rangle=\sum_{\mu\ell m} \psi_{\mu\ell m} |\mu\ell m\rangle\in\mathscr{H}.
\end{equation}
The left-hand side of the eigenequation
\begin{equation}
\label{eq:eigenvalue}
	\hat\theta \vert\psi\rangle = \vert\psi\rangle
\end{equation}
yields
\begin{align}
	\hat\theta \vert\psi\rangle
	=& \hat\theta \sum_{\mu\ell m} \psi_{\mu\ell m} |\mu\ell m\rangle
	= \sum_{\mu\ell m} \psi_{\mu\ell m}^\ast \text{e}^{\text{i}\theta_{\ell m}} \vert \mu\ell\;-\!m\rangle  
			\nonumber	\\
	=& \sum_{\mu\ell m} \left|\psi_{\mu\ell m}\right|
		\exp\left[-\text{i}\arg\psi_{\mu\ell m}+\text{i}\theta_{\ell m}\right] \vert \mu\ell\;-\!m\rangle,
\end{align}
and the right-hand side of~(\ref{eq:eigenvalue}) gives
\begin{align}
	\vert\psi\rangle
	=&  \sum_{\mu\ell m} \psi_{\mu\ell m} |\mu\ell m\rangle
	=  \sum_{\mu\ell m} \psi_{\mu\ell\;-\!m} \vert \mu\ell\, -m\rangle  
		\nonumber	\\
	=& \sum_{\mu\ell m} \left|\psi_{\mu\ell\;-\!m}\right|\exp\left[
		\text{i}\arg\psi_{\mu\ell\;-\!m}\right] | \mu\ell\;-\!m\rangle.
\end{align}
As the bases coincide, real and imaginary coefficients can be equated.
Normalization implies
\begin{equation}
\label{eq:|lambda|}
	\vert\psi_{\mu\ell\;-\!m}\vert=\vert\psi_{\mu\ell m}\vert.
\end{equation}

The imaginary part implies that
\begin{equation}
\label{eq:args}
	\arg\psi_{\mu\ell m} + \arg\psi_{\mu\ell\;-\!m} = \theta_{\ell m} \mod 2\pi.
\end{equation}
Notice that the left-hand side is symmetric under a change of the sign of $m$, 
which is true for integer~$m$ when~$\theta_{\ell m}$ varies as~$\pm\pi m$.

From conditions~(\ref{eq:|lambda|}) and (\ref{eq:args}), the eigenvectors of~$\hat\theta$ are
given by Eq.~(\ref{eq:eigenbasis}).
\end{proof}

\begin{remark}
	If one were to consider spinor representations of SO(3),
	then both~$\ell$ and~$m$ would be half-odd integer. 
	In this case the condition on~$m$ in~(\ref{eq:args}) fails, 
	which implies that every spinor state is a TR resource.
	\end{remark}

\begin{definition}
	A \emph{TR-invariant pure state} is a vector $|\psi\rangle\in\mathscr{H}$
	whose corresponding projection is equal to its TR group average
	\begin{equation}
	\label{eq:IsEig}
		\vert\psi\rangle\langle\psi\vert
		= \frac{1}{\vert S_2\vert} \sum_{g\in S_2}\Theta(g) \vert\psi\rangle\langle\psi\vert\Theta(g)^\dag
		= \frac{1}{2} \openone \vert\psi\rangle\langle\psi\vert \openone + \frac{1}{2}
			\hat\theta \vert\psi\rangle\langle\psi\vert \hat\theta^\dag
		= \frac{1}{2} \vert\psi\rangle\langle\psi\vert + \frac{1}{2} \vert\tilde{\psi}\rangle\langle\tilde{\psi}\vert, 
\end{equation}
for~$\vert\tilde{\psi}\rangle:=\hat\theta\vert\psi\rangle$.
\end{definition}
A state $|\psi\rangle$ is a TR resource if it is not invariant under TR, 
as it then must contains some informaton about time's `arrow'.
Thus, the states $|\mu\ell m\rangle$ are resources as they are not TR-invariant.
To begin studying TR resources, we first
identify zero-resource states. From~(\ref{eq:IsEig}), we see that these are necessarily eigenstates of~$\hat\theta$, and must satisfy $\hat\theta|\psi\rangle=e^{i\phi}|\psi\rangle$.
Hence, the states $|e_n\rangle$ are non-resource states.
Working with the self-conjugate basis~(\ref{eq:eigenbasis}) simplifies the behavior of pure states under time reversal:
\begin{equation}\label{eq:thetaaction}
\hat\theta \sum_{n} \psi_{n} |e_n\rangle 
	= \sum_{n} \psi^{\ast}_{n} |e_n\rangle.
\end{equation}
This leads us to the complete characterization of non-resource states:
\begin{lemma}
The state $|\psi\rangle$ is TR invariant state (i.e. non-resource state) iff $\psi_n = \langle e_n|\psi\rangle$ is {\it real} for all $n$.
\end{lemma}
\begin{proof}
A state $|\psi\rangle$ is TR invariant iff $|\psi\rangle=e^{i\phi}\hat\theta|\psi\rangle$.  In the self-conjugate basis of Eq.(\ref{eq:thetaaction}), this implies $\psi_n=e^{i\phi}\psi_n^\ast$.  Letting $\psi_n=|\psi_n|e^{i\phi_n}$, we must have $\phi_n=\phi/2$ mod$\pi$.  So $\psi_n = |\psi_n|e^{i\phi/2}$, which is real up to an unimportant $n$-independent global phase.
\end{proof}

\section{Time reversal invariant operations}

If two parties, Alice and Bob, lack a shared reference frame that informs each party of the
other's direction of time, their descriptions of the other's operators requires averaging over
the group action that relates their frames. Schur's Lemma tells us that this averaging
results in group invariant (or covariant) operators~\cite{BRS07}.  
In this section we classify the completely positive (CP) maps that are 
Time Reversal Invariant Operations (TRIO).

\subsection{Efficient maps}
We begin with a study of efficient maps.
\begin{definition}
	An \emph{efficient map} is a map whose Kraus decomposition comprises only one Kraus operator.
\end{definition}
Unitary (reversible) operations are examples of efficient CP maps.
The Kraus operator can be expressed in the $|e_n\rangle$ representation as
\begin{equation}
\label{eq:kraus}
	K=\sum_{nn'} K_{nn'}
		|e_n\rangle\langle e_n'|.
\end{equation}
The superoperator
\begin{equation}
\label{eq:K}
	\mathcal{K}(\bullet)=K \bullet K^{\dag}
\end{equation}
acting on a state projector~$\bullet\rightarrow|\psi\rangle\langle\psi|$ is a TRIO if
\begin{equation}
\label{eq:cond}
	K\hat{\theta}|\psi\rangle\langle\psi|\hat{\theta}^{\dag}K^{\dag}
		=\hat{\theta}K|\psi\rangle\langle\psi|K^{\dag}\hat{\theta}^{\dag}
		\;\forall |\psi\rangle\in \mathscr{H}.
\end{equation}

\begin{theorem}
\label{theorem:ABCD}
	An efficient map $\mathcal{K}(\bullet)=K \bullet K^{\dag}$ is TRIO iff $K_{nn'}$ are all real (that is, $K$ is real in the self-conjugate basis).
\end{theorem}

\begin{remark}
\label{remark:constantglobalphase}
	Note that a constant global phase~$\varphi$ can always be eliminated because
	\begin{equation}
	\label{eq:KM}
		\mathcal{K}(\bullet)=K \bullet K^{\dag}=(\text{e}^{\text{i}\varphi}K)(\bullet)(\text{e}^{\text{i}\varphi}K)^{\dag}.
	\end{equation}
\end{remark}

\begin{proof}
Working with the self-conjugate basis, $K|\psi\rangle$ is real for any real $|\psi\rangle$ iff $K$ is itself real.
\end{proof}

\subsection{Standard resources}
\label{subsec:standardresources}

In this subsection we show that all TRIO resource states can be parameterized by a single
real variable~$\theta$.

\begin{theorem}
\label{theorem:theta}
	For all $|\psi\rangle\in\mathcal{H}$ 
	$\exists$ a unitary TRIO Kraus operator $K$ such that
\begin{equation}
K|\psi\rangle=\frac{1}{\sqrt{2}}\begin{pmatrix}
                      1 \\ 
                      \text{e}^{\text{i}\theta} \\ 
                      0 \\ 
                      \cdot\\
                      \cdot\\
                      0
              \end{pmatrix}, \; 0\leq\theta \leq\frac{\pi}{2}.
              \end{equation}
\end{theorem}
\begin{proof}
A unitary TRIO $K$ is real in the self-conjugate basis and therefore orthogonal.
Any state in this basis can be written as
\begin{equation}
|\psi\rangle
= \sum_n \psi_n |e_n\rangle
= \sum_n \left(\psi^\mathrm{R}_n + i \psi^\mathrm{I}_n\right) |e_n\rangle
\end{equation}
where the real numbers $\psi^\mathrm{R}_n$ and $\psi^\mathrm{I}_n$ define real vectors $\psi^\mathrm{R}$ and $\psi^\mathrm{I}$.
Let $a=\|\psi^R\|$, and define $x = \psi^R/a$, a normalised real vector. 
Further, $\psi^{I}$ can be decomposed as $\psi^{I} = bx + cy$, where $y$ is a normalized real vector orthogonal to $x$. Thus we 
have $\psi = (a + ib)x + icy$, with a normalization condition $a^2 + b^2 + c^2 = 1$. Now let $x$ and $y$ be the two first columns of a unitary $K'$, then 
$\psi = K'(a + ib, ic, 0, . . . , 0)^T$. 

By a further $2\times2$ rotation $R_\alpha$, governed by a single rotation angle $\alpha$, $\psi$ can be brought to the form  
$|\psi\rangle=e^{i\gamma}K' R_\alpha \left(1, e^{\text{i}\theta},0,...,0\right)^T\equiv K\left(1, e^{\text{i}\theta},0,...,0\right)^T$, where $\gamma$ is an irrelevant global phase. 
To find $\alpha$ one has to look at the Bloch sphere 
representations of the density matrices of $(a + ib, ic)^T$ and $(1, e^{i\theta} )^T /\sqrt{2}$. Since both are pure states, they lie on the 
surface. The latter state is a point in the equatorial plane. As a real rotation has the effect of rotating along an axis in 
that same plane (to wit, the axis corresponding to the $\sigma_y$ Pauli matrix, which is left invariant under real rotations), any 
point on the sphere can be covered. 
\end{proof}
Theorem~\ref{theorem:theta} shows that all
TRIO resources are quantified by a single real variable $\theta$.

\subsection{Monotones and Interconvertibility}
\label{subsec:monotonesinterconvertibility}

From Theorem~\ref{theorem:theta}, every resource can be characterized by a qubit state of the form
\begin{equation}
\label{eq:standardstate}
	|\psi\rangle
		=\frac{\text{e}^{\text{i}\theta/2}}{\sqrt{2}}\left(|0\rangle+\text{e}^{-\text{i}\theta}|1\rangle\right)
		=\frac{1}{\sqrt{2}}\left(\text{e}^{\text{i}\theta/2}|0\rangle+\text{e}^{-\text{i}\theta/2}|1\rangle\right)
\end{equation}
where $0\leq\theta\leq\pi/2$.
We refer to Expression~(\ref{eq:standardstate})  as the standard representation of the resource.
From theorem~\ref{theorem:ABCD}, the allowed Kraus operators (i.e.\ TRIO invariant efficient operations) are therefore 
$2\times 2$ \emph{real} matrices, and the following definition of a monotone is thus reasonable.
\begin{definition}
	For $|\psi\rangle=\psi_0|0\rangle+\psi_1|1\rangle$ with $|\psi_0|^2+|\psi_1|^2=1$,
	the \emph{time-reversal monotone}~$\tau\left(|\psi\rangle\right)$ is
	\begin{equation}
		\tau\left(|\psi\rangle\right)=1-|\langle \psi^*|\psi\rangle|=1-|\psi_0^2+\psi_1^2|=1-\cos\theta
	\end{equation}
\end{definition}

\begin{theorem}
The TR monotone~$\tau$ is an ensemble monotone under TRIO.
\end{theorem}
\begin{proof}
Consider a measurement represented by TRIO Kraus operators $K_k$
that transforms the initial pure state to the ensemble
\begin{equation}
\label{eq:Kk}
	|\psi\rangle\mapsto\left\{p_k,\;|\varphi_k\rangle\equiv\frac{1}{\sqrt{p_k}}K_k|\psi\rangle\right\},
\end{equation}
where $p_k$ is the probability of each outcome.
Thus,
\begin{equation}
\sum_kp_k\tau(|\varphi_k\rangle) 
	= \sum_kp_k\left(1-\left|\langle\varphi_k^*|\varphi_k\rangle\right|\right)
	= 1-\sum_kp_k \left|\langle\varphi_k^*|\varphi_k\rangle\right|
	= 1-\sum_k \left|\langle(K_k\psi)^*|K_k\psi\rangle\right|,
\end{equation}
which leads to the inequality
\begin{equation}
	\sum_kp_k\tau(|\varphi_k\rangle)\leq 1-\left|\sum_k \langle(K_k\psi)^*|K_k\psi\rangle\right|
		= 1-\left|\sum_k \langle\psi^*|K_k^{\dag}K_k|\psi\rangle\right|
		= 1-|\langle \psi^*|\psi\rangle|=\tau\left(|\psi\rangle\right),
\end{equation}
where we have used the fact that $K_k$ are real matrices and 
$\sum_kK_k^{\dag}K_k=\openone$.
\end{proof}

\begin{theorem}
\label{theorem:Nielsenanalogue}
(Analogue of Nielsen's theorem~\cite{Nie99})
The transformation $|\psi\rangle\rightarrow|\varphi\rangle$ can be achieved 
deterministically by TRIO iff
$\tau\left(|\psi\rangle\right)\geq\tau\left(|\varphi\rangle\right)$.
\end{theorem}
\begin{proof}
As~$\tau$ is a monotone, the transformation is achievable by TRIO
provided that the condition~$\tau\left(|\psi\rangle\right)\geq\tau\left(|\varphi\rangle\right)$ is met.
Thus the condition is necessary, and now we have to show that the condition is also sufficient.

To show sufficiency, observe that, without loss of generality,
$|\psi\rangle$ and $|\varphi\rangle$ are expressed in our standard form
\begin{equation}
	|\psi\rangle=\frac{1}{\sqrt{2}}\left(\text{e}^{\text{i}\theta/2}|0\rangle+\text{e}^{-\text{i}\theta/2}|1\rangle\right)
\end{equation}
and
\begin{equation}
	|\varphi\rangle
		=\frac{1}{\sqrt{2}}\left(\text{e}^{\text{i}\gamma/2}|0\rangle+\text{e}^{-\text{i}\gamma/2}|1\rangle\right),
\end{equation}
where~$\theta,\;\gamma\in[0,\pi/2]$. The condition 
$\tau\left(|\psi\rangle\right)\geq\tau\left(|\varphi\rangle\right)$ is therefore equivalent to 
$\cos\gamma\geq\cos\theta$ (or~$\gamma\leq\theta$). 
We define a measurement in terms of its two Kraus operators,
\begin{equation}
	\hat{K}_1
		=\begin{bmatrix} 
			\sqrt{A/2} & \sqrt{(1-A)/2} \\ 
			\sqrt{A/2} & -\sqrt{(1-A)/2} 
		\end{bmatrix}
\end{equation}
and
\begin{equation}
	\hat{K}_2
		=\begin{bmatrix} 
			-\sqrt{(1-A)/2} & \sqrt{A/2}\\ 
			\sqrt{(1-A)/2} & \sqrt{A/2}
		\end{bmatrix},
\end{equation}
with
\begin{equation}\label{eq:A}
A = \frac{1}{2}+\frac{1}{2}\sqrt{\frac{\cos^2\gamma-\cos^2\theta}{1-\cos^2\theta}}.
\end{equation}
(Note that~$\theta=0 \implies \gamma=0$ and the problem is trivial.)  As the Kraus operators are real, the measurement is TRIO. 

By Luder's update rule, subsequent to the measurement, for this choice of the parameter $A$, 
the state is either
\begin{equation}
	|\varphi_1\rangle=\sqrt{2}\hat{K}_1|\psi\rangle
\end{equation}
or
\begin{equation}
	|\varphi_2\rangle=\sqrt{2}\hat{K}_2|\psi\rangle
\end{equation}
with~$$\tau(|\varphi_1\rangle)=\tau(|\varphi_2\rangle)=\tau\left(|\varphi\rangle\right).$$
Thus, $|\varphi_1\rangle,\;|\varphi_2\rangle$ and $|\varphi\rangle$ are equivalent up to
an orthogonal matrix corresponding to a reversible TRIO.
\end{proof}

\begin{theorem}(Analogue of Jonathan-Plenio theorem~\cite{JP99})
\label{theorem:JonathanPlenioanalogue}
Every transformation~$\mathcal{T}:
\left\vert \psi \right\rangle \rightarrow \{p_k,\left\vert 
\varphi_k\right\rangle \}$ that does not increase~$\tau$ on average,
i.e.\ for which
\begin{equation}
\label{eq:1a1}
	\sum_kp_k\tau(|\varphi_k\rangle )\leq \tau (|\psi \rangle ),
\end{equation}
can be achieved by some TRIO.
\end{theorem}

\begin{proof}
Without loss of generality, we assume the states to be in the standard form
\begin{equation}
	|\psi\rangle
		=\frac{1}{\sqrt{2}}\left(\text{e}^{\text{i}\theta/2}|0\rangle+\text{e}^{-\text{i}\theta/2}|1\rangle\right)
\end{equation}
and
\begin{equation}
	|\varphi_k\rangle=\frac{1}{\sqrt{2}}\left(\text{e}^{\text{i}\gamma_k/2}|0\rangle+\text{e}^{-\text{i}\gamma_k/2}|1\rangle\right),
\end{equation}
with~$\theta,\;\gamma_k\in[0,\pi/2]$.

We now define the state
\begin{equation}
	|\bar{\varphi}\rangle \equiv \frac{1}{\sqrt{2}}\left(\text{e}^{\text{i}\gamma/2}|0\rangle+\text{e}^{-\text{i}\gamma/2}|1\rangle\right) ,
\label{eq:varphibar}
\end{equation}
where~$\gamma$ is defined by
\begin{equation}
	\cos\gamma\equiv \sum_kp_k\cos\gamma_k.  \label{eq:t1}
\end{equation}
Noting that
\begin{equation}
	\tau(|\varphi _k\rangle )=1-\cos\gamma_k
\end{equation}
and
\begin{equation}
	\tau\left(|\bar{\varphi}\rangle\right)
		=1-\cos\gamma,
\end{equation}
we infer from Eq.~(\ref{eq:t1}) that
\begin{equation}
	\tau(|\bar{\varphi}\rangle )=\sum_kp_k\tau(|\varphi_k\rangle ).
\end{equation}
From Eq.~(\ref{eq:1a1}) we obtain
$\tau(|\bar{\varphi} \rangle )\leq \tau(|\psi \rangle)$, which implies, 
by Theorem~\ref{theorem:Nielsenanalogue},
that the transformation $|\psi\rangle\mapsto|\bar{\varphi}\rangle $ is achievable deterministically by TRIO.
Therefore, we only need to show that we can
generate the ensemble~$\{(p_k,|\varphi_k\rangle )\}$
starting from $|\bar{\varphi}\rangle$.

For this purpose, we define a set of Kraus operators
\begin{equation}
	\left\{K_k
		\equiv\begin{pmatrix}
			a_k&b_k \\ b_k&a_k
		\end{pmatrix}\right\},
\end{equation}
where
\begin{equation}
	a_k=\frac{1}{2}\sqrt{p_k}\left[\frac{\cos(\gamma_k /2)}{\cos(\gamma/2)}+\frac{\sin(\gamma_k/2)}{\sin(\gamma/2)}\right]
\end{equation}
and
\begin{equation}
	b_k 
		=\frac{1}{2}\sqrt{p_k}\left[\frac{\cos(\gamma_k /2)}
			{\cos(\gamma/2)}-\frac{\sin(\gamma_k/2)}{\sin(\gamma/2)}\right].
\end{equation}

Then it is straightforward to show that Relation~(\ref{eq:t1}) implies that 
\begin{equation}
	\sum_kK_k^{\dag }K_k=\openone,	\;\;
	K_k|\bar{\varphi}\rangle =\sqrt{p_k}|\varphi_k\rangle .
\end{equation}
Thus, the combination of this measurement with the deterministic protocol 
$|\psi \rangle \rightarrow |\bar{\varphi}\rangle $ realizes the required
transformation~$\mathcal{T}$. 
\end{proof}

\begin{corollary} (Analogue of Vidal's theorem~\cite{Vid99})
The maximum probability to convert $|\psi\rangle$ to $|\varphi\rangle$ by TRIO 
is given by
\begin{equation}
	P_\text{max}\left(|\psi\rangle\mapsto|\varphi\rangle\right)
		=\min\left\{\frac{\tau\left(|\psi\rangle\right)}{\tau\left(|\varphi\rangle\right)},1\right\}
\end{equation}
\end{corollary}

\begin{proof}
Consider a transformation that takes $|\psi\rangle$ to $|\varphi\rangle$ with probability $p$ and
takes $|\psi\rangle$ to $|0\rangle$ with probability $1-p$. 
As~$\tau(|0\rangle)=0$ (since it is an eigenstate of~$\hat\theta$, albeit not in standard form), Theorem~\ref{theorem:JonathanPlenioanalogue}
implies that such a TRIO transformation exists iff
\begin{equation}
	p \leq \frac{\tau\left(|\psi\rangle\right)}{\tau\left(|\varphi\rangle\right)}.
\end{equation}
\end{proof}

This corollary provides an operational interpretation of the measure~$\tau$.

\section{The Asymptotic Limit}
\label{sec:asymptotic}

In the previous section, the aim has been to determine whether a state is convertible to
another state and what resources are required for this conversion.
In this section we consider converting a state and its copies to another state plus copies.

\begin{definition}
	A \emph{copy} of a state~$|\psi\rangle\in\mathscr{H}$ is a state $|\psi'\rangle$ in
	a different Hilbert space $\mathscr{H}'$ such that the description of~$|\psi'\rangle$
	is identical to the description of~$|\psi\rangle$. The state with its copy is written
	$|\psi\rangle\otimes|\psi\rangle$, and~$n$ copies of the state are expressed as
	$|\psi\rangle^{\otimes n}$.
\end{definition}

Specifically this section concerns transformations of the type
\begin{equation}
	|\psi\rangle^{\otimes n}\rightarrow |\varphi\rangle^{\otimes m}
\end{equation}
with the integers $n$ and $m$ allowed to tend to infinity.
As discussed in Sec.~\ref{sec:representations}, the phase convention~(\ref{eq:LLconvention}) respects composition of systems via the decomposition of tensor product states into total angular momenta.
Here we will use such decompositions to show that arbitrary tensor powers of standard states can be brought to standard form with TRIO operations, i.e.
\begin{equation}
\label{eq:thetan}
	|\psi\rangle^{\otimes n}
		=  \left[ \frac{1}{\sqrt{2}} \left( \text{e}^{\text{i}\theta/2}|0\rangle + \text{e}^{-\text{i}\theta/2}|1\rangle \right) \right]^{\otimes n} 
	=:  \frac{1}{\sqrt{2}}\left(\text{e}^{\text{i}\theta_n/2}|0\rangle+\text{e}^{-\text{i}\theta_n/2}|1\rangle\right).
\end{equation}
We will give~$\theta_n$ in terms of the original~$\theta$ and finally define a frameness monotone that gives the maximum amount of distillable frameness from an ensemble.

It is not difficult to see that an arbitrary tensor product of standard states (\ref{eq:thetan}) can be written as a binomial distribution
\begin{equation}\label{eq:binom}
|\psi\rangle^{\otimes n} = \sum_{k=0}^n \sqrt{r_k} |\phi_k\rangle, \quad r_k = \frac{1}{2^n} \binom{n}{k} \text{e}^{\text{i}\theta(n-2k)},
\end{equation}
where $|\phi_k\rangle$ is the (normalised) equal superposition of all $n$-fold tensor products 
with $k$ standard $|1\rangle$ states and $n-k$ standard $|0\rangle$ states
\begin{equation}
	|0\rangle :=  |\mu=1,\ell=0,m=0,\epsilon=+\rangle,
	|1\rangle :=  |\mu=1,\ell=1,m=0,\epsilon=+\rangle.
\end{equation}
When decomposed, the states $|\phi_k\rangle$ will therefore contain total angular momenta 
up to~$\ell=k$ with $m=0$.

From Eqs.~(\ref{eq:eigenbasis}) and~(\ref{eq:LLconvention}), we can write such states as
\begin{equation}
|\mu\ell 0 +\rangle = \text{i}^\ell|\mu\ell m\rangle.
\end{equation}
Combining this with standard Clebsch-Gordan identities, we arrive at the following two decompositions that allow us to write any $|\phi_k\rangle$ in the preferred basis (\ref{eq:discretewavefunction})
\begin{align}
	|\mu \ell 0 +\rangle \otimes |0\rangle = & |\mu' \ell 0 + \rangle \\
	|\mu \ell 0 +\rangle \otimes |1\rangle = & \sqrt{\frac{\ell}{2\ell+1}}|\mu',\ell-1, 0 +\rangle
                                 + \sqrt{\frac{\ell+1}{2\ell+1}}|\mu'', \ell+1, 0 +\rangle;
\end{align}
both are symmetric in the tensor factors, and normalisation is of course preserved.
All angular momenta resulting from such coupling are given their own multiplicity index
because the resulting states must be orthogonal to any states previously occuring in a decomposition.
Theorem~\ref{theorem:theta} allows us to write the resulting state in the standard form (\ref{eq:thetan}).

Moreover, because 
\begin{equation}
	\sum_{k=0}^{n}r_k=\cos^n\theta
\end{equation}
is invariant under orthogonal transformations, by comparing the real norms of~(\ref{eq:thetan}) and~(\ref{eq:binom}), we find
that
\begin{equation}
\cos\theta_n=\cos^n\theta.
\end{equation}
This last observation motivates us to propose the following definition
of an operational measure for the asymptotic TR resource.

\begin{definition}
	Let $|\psi\rangle$ be a TR resource. We define
	\begin{align}
		\tau^{\infty}\left(|\psi\rangle\right)\equiv & -\log\left[1-\tau\left(|\psi\rangle\right)\right]
				\nonumber	\\
			=& -\log\left|\langle\psi^*|\psi\rangle\right|
				\nonumber	\\
			=& -\log\cos\theta.
	\end{align}
\end{definition}
\begin{lemma}
The measure~$\tau^\infty$ is additive; that is,
\begin{equation}
\tau^{\infty}(|\psi\rangle|\phi\rangle)=\tau^{\infty}\left(|\psi\rangle\right)+\tau^{\infty}(|\phi\rangle).
\end{equation}
\end{lemma}
\begin{proof}
\begin{align}
	\tau^{\infty}(|\psi\rangle|\phi\rangle)
		=&-\log\left|\langle\psi^*|\psi\rangle\langle\phi^*|\phi\rangle\right|
			\nonumber	\\
		=&\tau^{\infty}\left(|\psi\rangle\right)+\tau^{\infty}(|\phi\rangle).
\end{align}
\end{proof}

\begin{theorem}
	Given two resources $|\psi\rangle$ and $|\phi\rangle$ and an integer $n$, let the integer $m$
	be the maximum integer such that the transformation 
	$|\psi\rangle^{\otimes n}\rightarrow|\phi\rangle^{\otimes m}$
	is possible by TRIO. Then,
	\begin{equation}
		\lim_{n\rightarrow\infty}\frac{m}{n}
			=\frac{\tau^{\infty}\left(|\psi\rangle\right)}{\tau^{\infty}(|\phi\rangle)}.
	\end{equation}
\end{theorem}
\begin{proof}
As~$\tau$ is a frameness monotone and as~$\tau^\infty$ is a monotonic function of~$\tau$,
it follows that~$\tau^\infty$ is also a monotone. Hence, the transformation 
$|\psi\rangle^{\otimes n}\rightarrow|\phi\rangle^{\otimes m}$
is possible by TRIO if and only if~$\tau^\infty(|\psi\rangle^{\otimes n})\geq
\tau^\infty(|\phi\rangle^{\otimes m})$. This last inequality is equivalent to
\begin{equation}
\frac{m}{n}\leq\frac{\tau^{\infty}\left(|\psi\rangle\right)}{\tau^{\infty}(|\phi\rangle)}.
\end{equation}
Equality can obviously be achieved in the limit $n\rightarrow\infty$.
\end{proof}

Note that even though~$\tau^\infty$ is a monotone, it is \emph{not} an ensemble monotone.
This also happened with the chirality monotone in~\cite{GS08}.

\section{Conclusions}

We have addressed the problem of breaking time-reversal (TR) superselection by sharing a TR resource. 
Although TR invariance seems exotic because of its relevance to the counter-intuitive problem
of two parties trying to communicate but not knowing they are traveling the same or
opposite directions in time, time-reversal is pertinent to the more practical question of how
to encode quantum information into states of motion when a shared motional reference
is a consumable resource. 

We solve the problem of finding quantum resources for breaking TR superselection by
introducing TR-invariant operations, which we call TRIOs, and thereby constructing TRIO
equivalence classes. From these equivalence classes, we construct a resource theory
for overcoming the superselection rule. The resource theory leads to three types of 
frameness monotones depending on whether the TRIO process should be
deterministic, ensemble, or stochastic. 

Whereas studies of reference frame resources have so far involved unitary representations,
TR resources involve an anti-unitary representation; thus our treatment broadens the 
scope of reference frame resource theories.

\section*{Acknowledgements}
We appreciate many discussions with Rob Spekkens that inspired and initiated this work. 
We also appreciate valuable discussions with T.\ Rudolph and N. R. Wallach; and financial support from
the Alberta Ingenuity Fund, 
Alberta's Informatics Circle of Research Excellence (\emph{i}CORE),
Canada's Natural Sciences and Engineering Research Council (NSERC), 
the Canadian Network Centres of Excellence for Mathematics of Information
Technology and Complex Systems (MITACS), and General Dynamics Canada.
BCS is an Associate of the Canadian Institute for Advanced Research (CIFAR).

\end{document}